\documentstyle[aas2pp4]{article}

\def\as{$^{\prime\prime}$} 
 
\def\bdm{\begin{displaymath}} 
\def\edm{\end{displaymath}} 
\def\beq{\begin{equation}} 
\def\eeq{\end{equation}} 
\def\bit{\begin{itemize}} 
\def\eit{\end{itemize}} 
\def\ben{\begin{enumerate}} 
\def\een{\end{enumerate}} 
\def\bfi{\begin{figure}[htb]} 
\def\bpfi{\begin{figure}[p]}
\def\mum{$\mu m$}

\def\ea{{\it et al.}}

\def\paa{$\rm Pa\,\alpha$}
\def\sbpaa{$\rm SB_{Pa\alpha}$}
\def\ha{$\rm H\,\alpha$}
\def\h2o{$\rm H_2O$}

\begin{document}

\title{The NICMOS snapshot survey of nearby galaxies}

\received{Nov. 3, 1998}
\accepted{March 19, 1999}
%\journalid{}{}
%\articleid{}{}

\author{T. B\"oker\altaffilmark{1}, D. Calzetti, W. Sparks, 
D. Axon\altaffilmark{1}, L. E. Bergeron, 
H. Bushouse, L. Colina\altaffilmark{1}, D. Daou,
D. Gilmore, S. Holfeltz, J. MacKenty, L. Mazzuca, B. Monroe,
J. Najita, K. Noll, A. Nota\altaffilmark{1}, C. Ritchie, A. Schultz, 
M. Sosey, A. Storrs, A. Suchkov 
(the STScI NICMOS group)} 
\affil{Space Telescope Science Institute, 3700 San Martin Drive, 
Baltimore, MD 21218, U.S.A.}
\altaffiltext{1}{Affiliated with the Astrophysics Division, Space Science Department,
European Space Agency} 
\authoremail{boeker@stsci.edu}

\begin{abstract} 
We present ``snapshot'' observations with the
NearInfrared Camera and MultiObject Spectrometer (NICMOS) on board the
Hubble Space Telescope (HST) of 94 nearby galaxies from the Revised 
Shapley Ames Catalog.  Images with 0.2\as\ resolution were obtained 
in two filters, a broad-band 
continuum filter (F160W, roughly equivalent to the H-band) 
and a narrow band filter centered on the \paa\ line (F187N or F190N, 
depending on the galaxy redshift) with the 51\as $\times$51\as\ 
field of view of the NICMOS camera 3. A first-order continuum subtraction 
is performed, and the resulting line maps and integrated \paa\ line 
fluxes are presented. A statistical analysis indicates that the average \paa\
surface brightness {\bf in the central regions} is 
highest in early-type (Sa-Sb) spirals.
\end{abstract}
%%%%%%%%%%%%%%%%%%%%%%%%%%%%%%%%%%%%%%%%%%%%%%%%%%%%%%%%%%%%%%%%%%%%%
%%%%%%%%%%%%%%%%%%%%%%%%%%%%%%%%%%%%%%%%%%%%%%%%%%%%%%%%%%%%%%%%%%%%%
\keywords{infrared:galaxies---infrared:ISM:lines and bands---galaxies:nuclei---galaxies:starburst---galaxies:statistics}
%%%%%%%%%%%%%%%%%%%%%%%%%%%%%%%%%%%%%%%%%%%%%%%%%%%%%%%%%%%%%%%%%%%%%
%%%%%%%%%%%%%%%%%%%%%%%%%%%%%%%%%%%%%%%%%%%%%%%%%%%%%%%%%%%%%%%%%%%%%
\section{Introduction}\label{intro}
NICMOS is a second generation HST instrument, installed during the
HST servicing mission in February 1997. It extends the wavelength region
accessible for imaging and spectroscopy with HST to the near infrared (NIR),
up to $\lambda$ = 2.5~\mum . 

Shortly after its on-orbit installation, it was discovered that the
NICMOS dewar suffered from a thermal anomaly that led to a higher than
expected sublimation rate of the solid nitrogen coolant, and thus a shorter than
anticipated lifetime. In addition, the mechanical deformation of the dewar
prevents parfocality of the three NICMOS cameras, to the extent that NIC3,
the camera with the widest field of view (51\as$\times$51\as ), 
can only be brought fully to
focus by moving the secondary mirror of the HST. This, on the other hand,
prevents observations with the other HST instruments which is why it was
decided to concentrate all approved NIC3 observations in ``campaigns''
of 2-3~weeks duration.
Two of these campaigns have been successfully executed half a year
apart to allow all-sky access.

In order to fill any gaps in the observing schedule during the NIC3 campaigns
with meaningful scientific data, a ``snapshot'' survey was proposed 
and approved to be undertaken with Director's discretionary time 
(proposal ID 7919). The data were made public soon after the observations,
and can be retrieved from the HST archive.
The scientific rationale behind the project was to obtain wide field
imaging of the \paa\ emission for as many galaxies of the Revised Shapley Ames
Catalog (RSA, \cite{san87})
as allowed by the scheduling efficiency of the NIC3 campaigns.

Since the NIR is much less affected by dust extinction than optical 
wavelengths, the NIR is well suited for probing the heavily obscured
central regions of spiral galaxies.
The \paa\ line at 1.875~\mum\ is a recombination line of
atomic hydrogen, and as such is indicative of the ionized matter 
around hot, newly formed stars or active galactic nuclei (AGN).
It lies in a wavelength region subject to
strong atmospheric absorption by \h2o\ molecules and is therefore
accessible for ground-based observations only with great difficulty.
NICMOS provides two narrow band filters in the NIC3 camera 
that are suitable for \paa\ observations: the F187N filter was used for all
galaxies with a radial velocity $v_r$ between -1295 and 943~$km/s$, while
galaxies with 3181 $\leq v_r \leq 4780~km/s$ were observed with
the NIC3 F190N filter. The velocity range was chosen such that the
\paa\ line falls well inside the high transmission region of the respective
filter, with only minor corrections ($\leq$ 5\%) needed to account for
the highest velocities with respect to the filter central wavelength.

For continuum subtraction, each galaxy was also observed in the F160W
filter which approximately spans the H-band from 1.4 - 1.8~\mum .
%%%%%%%%%%%%%%%%%%%%%%%%%%%%%%%%%%%%%%%%%%%%%%%%%%%%%%%%%%%%%%%%%%%%%
%%%%%%%%%%%%%%%%%%%%%%%%%%%%%%%%%%%%%%%%%%%%%%%%%%%%%%%%%%%%%%%%%%%%%
\section{The sample} \label{sample}
The galaxy sample was selected at random from the RSA according to scheduling 
convenience. Ninety-four galaxies were observed in total, 64 of which 
lie in the low redshift range, and 30 in the higher. Sorted by Hubble type,
the sample contains 17 galaxies classified as E or S0, 20 Sa/Sab/Sb's, 
39 Sbc/Sc's, and 18 Scd/Sd/Sm/Irregulars. 
The sample thus should be large enough to be a fair
representation of the RSA and its distribution of Hubble type and luminosity.
Table \ref{tab1} lists the observed galaxies.

%Column~1 contains the NGC~number and possible alternative names.
%Columns~2 and 4 give the morphological type and recession velocity as
%taken from the RSA. Column~3 lists the type of nuclear activity
%for those sample galaxies that were classified by \cite{ho97a} (1) or
%\cite{ver98} (2). Galaxies that appear in neither of these catalogs
%have no entry. The integration time for the narrow band filter is listed in 
%column~5. Column~6 contains the total \paa\ flux in the field of view.
%Column~7 gives the ratio of the
%optical diameter $D_{25}$ as taken from the Second Reference Catalogue
%of Bright Galaxies (\cite{dev76}, RC2).
%Finally, Column~8 classifies the morphology of \paa\ emission,
%as explained in \S  \ref{results}. 

%--------------------
\begin{table}
\dummytable \label{tab1}
\end{table}
%--------------------
%%%%%%%%%%%%%%%%%%%%%%%%%%%%%%%%%%%%%%%%%%%%%%%%%%%%%%%%%%%%%%%%%%%%%
%%%%%%%%%%%%%%%%%%%%%%%%%%%%%%%%%%%%%%%%%%%%%%%%%%%%%%%%%%%%%%%%%%%%%
\section{Observations and data reduction} \label{data}
The observations were performed using the MULTIACCUM STEP64 sampling
sequence as
described in the NICMOS instrument handbook (\cite{mac97}). All F160W
continuum images were taken with NSAMP=11 for a total integration time
of 192~s. The F187N and F190N line emission images had values for NSAMP
between 13 and 25, yielding integration times between 320~s and 1088~s.
Column~5 of Table~\ref{tab1} contains the narrow band integration time.

For data reduction, the calibrated (*\_cal.fits) FITS images as 
retrieved from the HST archive were used as a starting point. These
images have been dark-subtracted, flatfielded, and cosmic ray corrected
in an automated way during the CALNICA pipeline processing as described in the 
HST data handbook (\cite{voi97}). The resulting images show a number of residual
anomalies, which are described in the following sections.
%%%%%%%%%%%%%%%%%%%%%%%%%%%%%%%%%%%%%%%%%%%%%%%%%%%%%%%%%%%%%%%%%%%%%
\subsection{Bad pixels}
Residual bad pixels in the CALNICA processed image are due to three possible 
defects: 
\ben
\item{Remaining cosmic ray hits that
are not detected by the CALNICA pipeline.}
\item{Defective (i.e. ``hot'' or ``dead'')
pixels on the NIC3 detector are also not replaced during
pipeline processing because the images were not dithered and no
meaningful observational data exist for these pixels.}
\item{Caused by mechanical contact due to the NICMOS dewar anomaly,
small flecks of anti-reflective paint have been scraped off one of the
optical baffles. Some of these have migrated onto the
NICMOS detector surfaces. These flecks, known as ``grot'', result
in small areas (one to a few pixels) of reduced sensitivity. Since
the grot might be subject to unpredictable movement on the detector, it is
currently not included in the bad pixel mask for CALNICA. However,
recent results have shown that the grot has been quite stable over time,
and plans exist to include the affected pixels in the next version of the 
CALNICA bad pixel masks.}
\een
Since the point-spread-function (PSF) is undersampled in
the NIC3 camera, it is not straightforward to discriminate between
uncorrected bad pixels and real point sources in the pipeline processed
data. We created a mask that defined as bad all pixels that showed
a deviation greater than 4$\sigma$ in at least 10\% of the images. 
Since real point sources (and - unfortunately - remaining cosmic
ray hits) should appear randomly in the field,
this ensures that only systematically bad pixels are included in the mask.
All masked pixels were then replaced with the median value of their neighboring
pixels. No attempt has been made to correct for the remaining cosmic rays. 
%%%%%%%%%%%%%%%%%%%%%%%%%%%%%%%%%%%%%%%%%%%%%%%%%%%%%%%%%%%%%%%%%%%%%
\subsection{Pedestal effect}
The so-called ``pedestal effect'' is a residual DC offset that appears
in an image after the dark current subtraction during CALNICA processing.
The dark current reference files that are used by CALNICA are not 
a perfect representation of the actual instrumental bias and dark current, 
because some or all of the components that contribute to the dark signal
are time variable. These variations are believed to be driven by
subtle temperature changes in the electronics and/or the detectors themselves.
Typical pedestal amplitudes are a few data numbers (DN) up to about
15~DN, but larger values have been occasionally observed. 

The effect of a residual bias in the image after dark current subtraction
is that multiplication with the flatfield reference file leaves an
imprint of the relative pixel sensitivities in the resultant image.
To further complicate the matter, the four quadrants of each NICMOS detector
have separate amplifiers and read-out electronics. Therefore 
the pedestal typically is different in each quadrant of a NICMOS image.

To correct the data for the pedestal effect, we have used
code developed at STScI by R. van der Marel. The code assumes a
constant bias in each detector quadrant, which is a good approximation in
most cases. In brief, the program determines the value of this bias
by minimizing the spread of the pixel values in each quadrant. The rationale 
behind this method is that any non-zero bias increases the spread 
because the image is multiplied by the (inverse) flatfield during 
CALNICA processing. The software and a description
of the algorithms used are available from the 
WWW\footnote{\it http://sol.stsci.edu/$\sim$marel/software/pedestal.html}.
For fields that are largely filled by the science object, as is the
case for the dataset presented here, the method of spread minimization 
performs better than other methods like
simple median subtraction or large-scale structure removal.

After the bias value has been determined for each quadrant, the
scaled flatfield is subtracted from the image, and the four quadrants
are brought to match. The last step is difficult to do in an automated
way, especially if the galaxy nucleus --- which typically has a very steep 
gradient in its signal --- is located close to quadrant borders. 
Any residual ``edges'' in the images presented in \S  \ref{results}
are due to shortcomings in the quadrant equilization step and can
be improved manually by adding small constants to each quadrant.
%%%%%%%%%%%%%%%%%%%%%%%%%%%%%%%%%%%%%%%%%%%%%%%%%%%%%%%%%%%%%%%%%%%%%
\subsection{Continuum subtraction} \label{cont_sub}
To subtract the continuum underlying the \paa\ line emission in the narrow 
band filter images, the following approach was taken:

Both the F160W and F187N/F190N images were rebinned to a size
of 128$\times$128 pixels for the purpose of reducing
both the scatter and the total number of datapoints in the plot.
We then plotted the intensity (in DN) of each
pixel in the rebinned F160W image versus that of the same pixel in the
rebinned F187N/F190N image. If the galaxy had constant color over
the field of  view and no \paa\ emission, the relation should be linear down to the
noise level. Deviations from the linear relation can be due to 
intrinsic color variations, differential extinction effects, \paa\ 
line emission, or a combination thereof. 

We then performed a linear least squares fit to the data points,
the slope of which should be the scaling factor between the two filters.
In most cases, all pixels were used for the fit, excluding the bottom
10~rows in the original images, since these
are subject to vignetting and contain no reliable information.
In addition, the 500 brightest
pixels in the F187N/F190N image were excluded, because those pixels typically
contain either foreground stars, bright HII regions, or the galaxy nucleus.
Therefore they do not provide a good estimate for the 
average color of the image. Finally, the F160W image was scaled by the 
slope of the linear fit, and the result subtracted from the F187N/F190N 
image in order to obtain a map of the excess emission. 
Although it is not possible to discriminate between the various contributions
to the excess emission, as mentioned above, we refer to the resulting
images as \paa\ maps for the remainder of this paper in order to simplify 
the terminology. We also point out that if a smoothly distributed
component of \paa\ emission is present, it will be removed by this
procedure, thus leading to an underestimate of the total \paa\ flux. 
Also, no attempt was made to correct for the PSF differences between the
two filters because the NIC3 PSF is undersampled in both filters. 

Fig. \ref{fig1} contains the F160W images,
the \paa\ maps, and the fit results for all galaxies of the sample
after the described data reduction.
The plot of the flux distribution gives a visual impression of the 
complexity of the galaxy color distribution or, correspondingly, 
the reliability of the continuum scaling factor. For most early type galaxies, 
e.g. NGC~2314 or NGC~2681, the scatter around the linear best fit is
small. In principle, a higher fraction of ionized gas causes larger scatter.
In nearby galaxies where many individual stars are resolved (e.g. NGC~247
or NGC~4144), the scatter is largest because of the color distribution of
the stars. However, a mere shift of the solid line
indicating the best fit does not affect the quantitative results. 
This is because a possible offset is removed when
measuring the \paa\ signal, as described in \S ~\ref{paa}

Based on the \paa\ images of Fig.~\ref{fig1},  
a rough morphological classification of the galaxy sample was performed. 
In column~8 of Table \ref{tab1}, each galaxy is labeled according to 
the presence of one or more of the following features:
unresolved or extended \paa\ emission from the galaxy nucleus (N), 
individual HII regions located in a spiral arm structure (S), a ring (R), 
or isolated throughout the disk (I), diffuse nebulosities (DN), 
or prominent dust lanes (DL). For most of these labels, we feel
confident that they indeed contain information on the distribution
of real \paa\ emission. The only exception is the often
observed signal from the nucleus of the galaxy. Whether this
is true \paa\ emission or the effect of a color gradient can only be
answered with additional spectral information. 
A slight spatial offset between the two images
or the small PSF differences between the filters
might also lead to residual asymmetric signal in the nuclei, 
if they have very steep brightness profiles. However, the edge-like color 
gradients observed in 
some nuclei like NGC~2683 or NGC~4373 are probably real, since 
stars in the field subtract out almost perfectly, and the structure is too
extended to be caused by PSF mismatches. 
In summary, the label N in Table \ref{tab1} indicates that the \paa\ image
shows increased signal in the nucleus without specifying its nature.

A number of objects show an obvious oversubtraction 
in the nucleus. This is a direct consequence of the simplistic ``one-color'' 
approach taken here if the continuum emission in the center has 
a bluer color than in the disk. The most likely explanations for a bluer color
of the nucleus are enhanced star star formation or the presence of an AGN.
A more detailed case-by-case study, possibly with
additional color information, should allow a better continuum subtraction
in the nuclei, but is beyond the scope of this paper due to the large
data volume. 
%%%%%%%%%%%%%%%%%%%%%%%%%%%%%%%%%%%%%%%%%%%%%%%%%%%%%%%%%%%%%%%%%%%%%
\subsection{\paa\ fluxes} \label{paa}
In order to compare the star formation properties of the galaxy sample,
the integrated \paa\ flux was calculated from the line maps.
A first step checked for any systematic offsets in the maps by
calculating the mode of pixel values in an empty region of the \paa\ map. 
The mode should be close to zero after the described data reduction procedure,
if the linear fit for the continuum was correct. 
In cases where a (small) offset was present, 
the mode was subtracted to bring the \paa\ ``background'' to zero. 

An automated method was developed to sum
the total \paa\ flux in the line maps which is briefly described here.
Continuum oversubtraction would seriously affect the result when 
simply adding all pixels in the \paa\ map. The image was therefore
clipped at a certain threshold and only the signal above this level was summed. 
One complication to this method is that some images 
have a number of pixels above the noise level that apparently do not contain 
true line emission. This is particularly true for images that were taken
after an HST passage through the South Atlantic Anomaly (SAA). The effect
of a cosmic particle hit on the NICMOS detectors is equivalent to a higher
dark current in the affected pixels. Therefore, when defining the threshold, 
we had to compromise between clipping
at a high level, thus potentially losing real \paa\ signal, or
including more noise in the number for \paa\ flux. For images that contained
extended \paa\ emission, we found that a 2~$\sigma$ threshold
would not lose a significant amount of \paa\ signal, while eliminating
most - but not all - of the noisy pixels. Therefore, all images
were clipped at the 2~$\sigma$ level, i.e. pixels below this threshold 
were set to zero, and the total remaining flux in the image calculated. 
A possible remaining noise contribution and the uncertainty 
regarding the signal nature described in \S  \ref{cont_sub} could lead to an
overestimate of the true \paa\ flux. On the other hand, any \paa\ contribution 
below the 2$\sigma$ level will be missed by our method, as well as a 
smooth, extended component.
 
Such a diffuse Pa-alpha emission is known to be present in at least some 
early type galaxies, and virtually all ellipticals and S0's for 
which optical spectra 
are available are classified as LINERs (\cite{gou98}). However, the exact 
physical mechanism behind the gas ionization is still unclear.
Several possibilities, including shocks, post-AGB stars, cooling flows, and a 
low luminosity AGN could be at work, but, contrary to spirals and irregulars, a 
recent star formation is discarded (\cite{gou98}). Thus, 
the removal of a possible diffuse component will not affect any conclusions
on star formation activity.

We estimate the uncertainty in the numbers 
of Table~\ref{tab1} to be between 10 and 30\%, depending on the noise
level. We made no attempt to
correct for the position of the line with respect to the
central wavelength of the narrow-band filter, because the required
corrections would not contribute significantly to the described uncertainties,
as explained in \S ~\ref{intro}.
%%%%%%%%%%%%%%%%%%%%%%%%%%%%%%%%%%%%%%%%%%%%%%%%%%%%%%%%%%%%%%%%%%%%%
%%%%%%%%%%%%%%%%%%%%%%%%%%%%%%%%%%%%%%%%%%%%%%%%%%%%%%%%%%%%%%%%%%%%%
\section{Results} \label{results}
This survey is an unbiased sample of galaxies closer than 
$v_z\,\sim\,5000~km/s$, and should therefore be representative of the local 
universe. In particular, the combination of narrow and broad band filters
reveals the \paa\ emission and should give us an unobscured view of 
star formation in the neighborhood of the Milky Way. 
Because of the large data volume, no attempt was made
to address spatial color gradients (see \S ~\ref{cont_sub}). 
However, a more detailed analysis on a case-by-case basis should be
possible, e.g. by comparing \paa\ equivalent widths with
predictions from stellar population synthesis models. A few general
comments can still be made from the images alone.
%%%%%%%%%%%%%%%%%%%%%%%%%%%%%%%%%%%%%%%%%%%%%%%%%%%%%%%%%%%%%%%%%%%%%
\subsection{Star formation in the galaxy disks}
From the images in Fig. \ref{fig1}, it is evident that \paa\ emission in
the disks of most spiral galaxies traces the spiral arm structure. 
This is expected since the denser environment in the spiral arms is
known to cause enhanced star formation. The line emission is clumpy,
unlike the generally smooth stellar continuum. This 
matches similar results obtained at ultraviolet
wavelengths which also trace active star formation (e.g. \cite{wal97}). 
However, the UV morphology is hampered by strong dust extinction.
The question is whether star formation is clumpy because of how
stars form or because of how dust obscures. In this context, the relatively
dust-insensitive \paa\ morphology in our images confirms earlier 
claims that star formation is intrinsically clumpy.  
Early type galaxies generally show little or
no line emission in their disks, but often have strong emission
from their central regions. 
A notable exception is NGC~1241, an Sb spiral with a Sy~2 nucleus, which shows 
an inclined ring of star formation  about 7\as\ in diameter 
surrounding the nucleus. At the distance of the galaxy (80~Mpc for 
H$_o$=50~km/s/Mpc), this corresponds to 2.7~kpc. 

The clumps of \paa\ emission lose their organized distribution when
going from late type spirals to irregular galaxies, and their typical
size becomes larger. The strongest \paa\ emission is found in 
starburst galaxies like NGC~3077, NGC~3593, or NGC~4449, to name a few. 
A special case is NGC~1705. This metal-poor, almost dust-free dwarf galaxy
contains a very UV bright super star cluster, which has very little
nebular gas emission (\cite{meu95}). This is an effect of the
cluster age of about 13~Myr. At this stage, the hottest cluster
stars are no longer ionizing the surrounding interstellar matter.
However, the strong stellar winds observed to come out of the
galaxy are believed to be a consequence of the past activity of the
cluster (\cite{hec97}). They might also be responsible for the
extended, low-level star formation in the outer regions of the galaxy that
is evident in the \paa\ image.

A number of highly inclined galaxies in the sample show strong dust 
lanes even in the NIR. Examples are NGC~891, 
and also NGC~2683 and NGC~5908. The dust lane in NGC~891 
is so opaque that the differential extinction in the two
filters causes strong residual signal in
the \paa\ map. This extreme case demonstrates the problem of disentangling
color effects from true line emission.
%%%%%%%%%%%%%%%%%%%%%%%%%%%%%%%%%%%%%%%%%%%%%%%%%%%%%%%%%%%%%%%%%%%%%
\subsection{\paa\ surface brightness and star formation rates}
It is customary to use the surface brightness (SB) of hydrogen recombination
lines, in particular the \ha\ line, as a measure for the star formation rate
of the galaxy. Studies along these lines at UV and optical wavelengths
(e.g. \cite{ken83a,ken83b,deh94}, Young \ea\ 1996)
resulted in the notion that star formation in later Hubble types is on the
average higher than in earlier types.

These studies were performed over apertures that
included the whole visible disk of the galaxy. In contrast, the NIC3 field of 
view of 51\as $\times$51\as\ in all cases does not contain the full disk,
especially for the nearby galaxies. This is evident from column~7
of Table~\ref{tab1} which lists the ratio $D_{25}$ to the NIC3 field of view.
Since it is unknown what fraction of the total \paa\ emission lies outside
the NIC3 field, it is not possible to derive the average \sbpaa\ over the
whole galaxy disk.
Therefore, one can only draw conclusions for the central 51\as\
of the galaxies, and thus might expect slightly
different results for the variation of star formation rate with Hubble type
than derived from the earlier studies. 
The \paa\ flux from Table~\ref{tab1} was converted into an average surface
brightness {\bf over the NIC3 field} according to 
%%%%%%%%%%%%%%%%%%
\beq \label{eq1}
\rm SB_{Pa\alpha} [L_{\odot}/pc^2] = 5\cdot 10^{18}\cdot F_{Pa\alpha} [W/cm^2] .
\eeq
%%%%%%%%%%%%%%%%%%
Here, the constant $5\cdot 10^{18}$ results purely from unit 
conversions, the mean value of \sbpaa\ is independent of galaxy distance.
Following the approach of \cite{you96}, each of the four panels in 
Fig.~\ref{fig2}a contains histograms of \sbpaa\ for the full
sample of 94 galaxies (dashed line) in comparison to a subset of
Hubble types, namely (i) E - S0, (ii) Sa - Sb, (iii) Sbc - Sc, and
(iv) Scd - Irr. The median \sbpaa\ for all galaxies in the 
respective subset
is indicated by a star symbol. As found by the earlier studies, the 
scatter in the histograms is large. We also confirm the generally low
star formation activity in Ellipticals and S0s, about an order of magnitude
fainter than the values found for spirals and irregulars. 

As for the spirals and irregulars in our sample, we do not see the 
monotonic increase
of \sbpaa\ between Sa and Sm that \cite{you96} found for \ha .
Rather, we find that \sbpaa\ {\bf decreases} slightly
between Sa and Sm. This result remains basically unchanged if 
all galaxies with redshifts above 945~$km/s$ (those observed with 
the F190N filter) are excluded (Fig~\ref{fig2}b). For the full sample,
Sa's to Sb's show a
median (mean) \sbpaa\ that is 53\% (64\%) higher than Sbc's and Sc's.

The images of our sample
are dominated by the central regions of the galaxies. This follows from the
fact that the average $D_{25}$ is larger than the field-of-view
(9.1 times the NIC3 field for the F187N sample, compared to 2.7 for 
the F190N sample, see column~7 in Table~\ref{tab1}). 
Therefore, our finding of a higher \sbpaa\ in the centers of
early-type galaxies can be understood if the fraction of ionized gas
in their nuclei is higher than in 
late types. A similar notion was made by \cite{you96}
who also found that a substantial fraction of the \ha\ emission in early
Hubble types stems from the nucleus. This result can be interpreted as a 
direct consequence of the more likely existence of an AGN in early-type spirals.
A large fraction of the early-type (Sbc or earlier) galaxies 
in our sample (15 out of 47) show indications of
a Seyfert- or LINER-type nucleus, while only a small fraction (2 out of 47) of 
late types (Sc or later) are classified as AGNs. Thus, if a compact
AGN exists preferentially in the core of early-type spirals, its contribution 
would increase the ionized gas fraction relative to a galaxy without an AGN.
Therefore, this result does not mean 
that nuclear star formation is weaker in late-type spirals but reflects 
the observational result that luminous AGNs avoid late spirals and irregulars, 
as found by \cite{ho97b}. In fact, many Sc's or Sd's reveal signs of massive 
gaseous inflow towards their nuclei, as will be shown in the next section.
%%%%%%%%%%%%%%%%%%%%%%%%%%%%%%%%%%%%%%%%%%%%%%%%%%%%%%%%%%%%%%%%%%%%%
\subsection{Nuclear star formation and gas dynamics}
Active star formation occurs in regions
where the molecular gas is dense enough to become gravitationally unstable.
The strong star formation that is often observed in the nuclear regions
of galaxies requires some mechanism that causes large amounts of molecular 
gas to fall into the central few hundred parsecs of the galaxy.

The most commonly evoked process by which the gas can lose its angular 
momentum is that of dynamical resonances of the gas orbits in the 
non-axisymmetric potential of a stellar bar. Various 
compelling theoretical arguments  
indicate that bars in galaxies are indeed an efficient mechanism for 
channeling gas from the outer to the nuclear regions of spirals
(\cite{shl90,ath92,but96}).  
According to these models, nuclear star-forming rings and disks are 
expected by-products of gas inflow towards the 
inner regions in barred spirals at the locations of the inner Lindblad 
resonances (ILRs). These nuclear rings are also thought to be an 
integral part of the gas dissipation processes which ultimately lead 
to the fueling of AGNs and possibly to the formation of central black holes.  
Massive nuclear disks are themselves subject to non-axisymmetric 
dynamical instabilities 
which drive the gas further inwards by means of gravitational
torques, possibly leading to the formation of nested bars and fueling
stellar and nonstellar activities in the center (\cite{shl89}). 

The unprecedented spatial resolution of the NICMOS data allows to
investigate such processes in more distant galaxies than
possible from ground-based observations. Figure~\ref{fig3} contains a 
collection
of some of the more interesting galaxies in our sample that reveal complex
gas distributions like double peaks, spirals, bars, rings or arcs in the
\paa\ images. All of them are Sbc or later spirals.
In Fig.~\ref{fig3}, we compare the \paa\ (in grayscale) 
and continuum (contours) emission from the nuclear region. 
To emphasize the extended structure, all images have been median filtered over
a 3$\times$3 box. In many cases in Fig.~\ref{fig3}, like, e.g., 
NGC~2989, NGC~6754,
or NGC~6946, the continuum contours are elongated. This is unlikely a
pure projection effect, and emphasizes that the prominent stellar bars 
found in many early-type spirals can indeed have weaker counterparts in 
late Hubble types. High spatial resolution NIR studies, together with
molecular line observations of a number of
nearby galaxies like e.g. M~83 (\cite{han90,gal91}), NGC~253 (\cite{pen96,boe98}), 
NGC~4570 (\cite{vdb98}), or IC~342 (\cite{ish90,boe97}), have
shown that massive gas inflows --- presumably triggered by dynamical resonances
with a stellar bar --- fuel active star formation in the central few
tens of parsecs. It is likely that the gas morphologies 
seen in Fig.~\ref{fig3} are caused by similar processes.
%%%%%%%%%%%%%%%%%%%%%%%%%%%%%%%%%%%%%%%%%%%%%%%%%%%%%%%%%%%%%%%%%%%%%
%%%%%%%%%%%%%%%%%%%%%%%%%%%%%%%%%%%%%%%%%%%%%%%%%%%%%%%%%%%%%%%%%%%%%
\section{Summary}
We have presented near infrared continuum and \paa\ images of the centers
of 94 nearby galaxies of all morphological types. The unprecedented spatial
resolution of the data reveals remarkable activity and a wealth of 
structure in many of the galaxies.

We have mapped the gas morphology in the galaxy nuclei and found in
many cases evidence for gas infall, most likely triggered by dynamical
interaction with a stellar bar. We also have performed a statistical
analysis of the average \paa\ surface brightness \sbpaa\ over the NIC3 field.
The main result is that \sbpaa\ 
{\bf in the central regions} is on the average stronger 
in Sa and Sb galaxies than in later Hubble types, most likely due of the
higher fraction of AGN in early type spirals.

The catalog is intended to serve as a database for further study of 
individual objects, their star formation activity, dynamics, and
matter distribution. The data are available electronically from
the HST archive (Prop. 7919), 
and we encourage all interested colleagues to take advantage of them.

We are grateful to our anonymous referee whose comments helped a great
deal to improve the quality of this paper.
This research has made use of the NASA/IPAC Extragalactic Database (NED) 
which is operated by the Jet Propulsion Laboratory, California Institute 
of Technology, under contract with the National Aeronautics and Space 
Administration. 
%%%%%%%%%%%%%%%%%%%%%%%%%%%%%%%%%%%%%%%%%%%%%%%%%%%%%%%%%%%%%%%%%%%%%
{}
%%%%%%%%%%%%%%%%%%%%%%%%%%%%%%%%%%%%%%%%%%%%%%%%%%%%%%%%%%%%%%%%%%%%%
%%%%%%%%%%%%%%%%%%%%%%%%%%%%%%%%%%%%%%%%%%%%%%%%%%%%%%%%%%%%%%%%%%%%%
\section*{Figure captions}
%----------------------
\figcaption[]{F160W image (top), \paa\ image (center), and results 
of the continuum fit (bottom)
for all galaxies of the sample. North and east are indicated 
in the upper right corner of the F160W image. The field of view
for all images is 51\as$\times$51\as , the resolution is 0.2\as . 
In the plots, the
100 brightest pixels in the F160W are not shown to optimize the 
plot range.  \label{fig1}}
%----------------------
\figcaption[]{Histograms of the \paa\ surface brightness for various
Hubble types. a) Results for the full sample. b) Results for
only those galaxies with redshifts less than 945~$km/s$. 
For each histogram, the results for the full
sample of 94 galaxies is indicated by the dashed line for comparison.
The median SB for all galaxies in the 
respective subset is indicated by a star symbol. \label{fig2} }
%----------------------
\figcaption[]{Overlay of the \paa\ (grayscale) and continuum (contours)
emission of the central regions for nine sample galaxies with central
gas concentrations. The images
have been median filtered over a 3$\times$3 box. \label{fig3} }
%----------------------

\newpage
\begin{deluxetable}{lccccccc}
\tablenum{1}
\tablewidth{0pt}
\tablecaption{The sample}
\tablehead{
\colhead{NGC} & \colhead{type} & \colhead{AGN?}& \colhead{$v_z$} 
	& \colhead{$\rm t_{int}$} & \colhead{F$_{\rm Pa\,\alpha}$} 
	& $\frac{D_{25}}{51^{\prime\prime}}$ & \colhead{\paa ?}  \\
 & & & [km/s] & [s] & [$10^{-21}$W/cm$^2$] & &  
	}         
\startdata
128 & S0(8)pec & & 4253 & 896* & 	$<$ 0.1 & 3.5 & -- \nl 
151 & SBbc(rs) & & 3741 & 704* & 	10.6 & 4.4 & R/S,N \nl 
214 & Sbc(r) & & 4499 & 704* & 		12.0 & 2.2 & S,N \nl 
221 (M~32) & E2 & & -200 & 832 & 	46.7 & 10.3 & -- \nl 
237 & Sc(s) & S?$^2$ & 4139 & 768* & 	18.5 & 1.9 & S \nl 
247 & Sc(s) & & 156 & 448 & 		8.2 & 25.3 & I \nl 
404 & S0(0) & L2$^1$ & -39 & 576 & 	16.8 & 4.1 & N \nl 
491 & SBbc(r) & & 3899 & 704* & 	43.8 & 1.7 & S \nl 
598 (M~33) & Sc(s) & H$^1$ & -180 & 896 & 	21.3 & 83.5 & DN \nl 
628 (M~74) & Sc(s) & & 656 & 576 & 	5.7 & 12.4 & I \nl 
672 (VV~338) & SBc(s) & H$^1$ & 413 & 1088 & 20.0 & 8.5 & I \nl 
891 & Sb & H$^1$ & 530 & 960 & 		147.2 & 15.9 & DL \nl 
925 & SBc(s) & H$^1$ & 564 & 576 & 	13.1 & 12.4 & I \nl 
976 & Sbc(r) & & 4362 & 512* & 		22.6 & 1.8 & S,N \nl 
1241 (VV~334) & SBbc(s) & S2$^2$ & 4028 & 576* & 	20.2 & 3.3 & S,R,N \nl 
1705 & Am. & & 640 & 576 & 		5.2 & 2.2 & I,DN \nl 
2314 & E3 & & 3872 & 704* & 		9.9 & 2.0 & N \nl 
2366 (DDO~42) & SBm & & 98 & 1088 & 	6.9 & 9.6 & I \nl 
2403 & Sc(s) & H$^1$ & 131 & 640 & 	20.0 & 25.8 & I, DN \nl 
2639 & Sa & S1.9$^2$ & 3187 & 640* & 	9.6 & 2.1 & R,N \nl 
2642 & SBb(rs) & & 4439 & 576* & 	9.8 & 2.4 & S,N \nl 
2672 & E2 & & 3983 & 384* & 		$<$ 0.1 & 3.5 & -- \nl 
2681 & Sa & L1.9$^1$ & 715 & 576 & 	24.3 & 4.2 & N \nl 
2683 & Sb & L2/S2$^1$ & 404 & 896 & 	32.9 & 11.0 & S,DL,N \nl 
2685 & S0(7)pec & S2/T2::$^1$ & 881 & 704 & 	2.7 & 5.3 & N \nl 
2749 & E3 & & 4229 & 896* & 		5.9 & 2.0 & N \nl 
2787 & SB0/a & L1.9$^1$ & 664 & 960 & 	$<$ 0.1 & 3.8 & -- \nl 
2841 & Sb & L2$^1$ & 637 & 768 & 	11.6 & 9.6 & N \nl 
2903 & Sc(s) & H$^1$ & 550 & 704 & 	74.5 & 14.9 & R,I,DN,DL\nl 
2942 & Sc(s) & & 4414 & 704* & 		12.7 & 2.6 & I \nl 
2976 & Sd & H$^1$ & 13 & 512 & 		14.4 & 7.0 & I,N \nl 
2989 & Sc(s) & H2$^2$ & 4166 & 960* & 	6.5 & 2.0 & S,N \nl 
2998 & Sc(rs) & & 4777 & 512* & 	7.7 & 3.4 & S,I \nl 
3077 & Am. & H$^1$ & 7 & 832 & 		66.5 & 6.4 & DN \nl 
3184 & Sc(r) & H$^1$ & 589 & 960 & 	6.4 & 8.7 & N,I \nl 
3271 (IC~2585) & Sa & & 3824 & 1088* & 	13.1 & 3.7 & N \nl 
3275 & SBab(r) & & 3241 & 640* & 	16.7 & 3.3 & N \nl 
3379 (M~105) & E0 & L2/T2::$^1$ & 893 & 704 & 	$<$ 0.1 & 6.4 & -- \nl 
3571 (NGC~3544) & Sa & & 3777 & 896* & 		$<$ 0.1 & 3.5 & -- \nl 
3593 & Sa~pec & H$^1$ & 625 & 320 & 	57.0 & 6.1 & S,R,N \nl 
3627 (M~66) & Sb(s) & T2/S2$^1$ & 723 & 640 & 	23.5 & 11.4 & N,I \nl 
3675 & Sb(r) & T2$^1$ & 762 & 832 & 	26.1 & 7.0 & S,I \nl 
3738 & Sd & H$^1$ & 224 & 896 & 	11.6 & 3.0 & I,DN \nl 
3769 & SBc(s) & & 737 & 512 & 		17.7 & 3.7 & I \nl 
3782 & SBcd(s) & & 738 & 832 & 		6.6 & 2.0 & I \nl 
IC~749 & SBc(rs) & & 798 & 896 & 	7.5 & 2.7 & I \nl 
IC~750 & S(b) & & 713 & 1088 & 		94.0 & 3.1 & S \nl 
4026 & S0(9) & & 878 & 768 & 		3.4 & 6.1 & DN \nl 
4062 & Sc(s) & H$^1$ & 772 & 576 & 	14.3 & 4.8 & I \nl 
4085 & Sc & & 753 & 768 & 		44.0 & 3.3 & S,N \nl 
4102 & Sb(r)~pec & H$^1$ & 865 & 704 & 	98.8 & 3.5 & S,N \nl 
4111 & S0(9) & H$^1$ & 791 & 896 & 	18.4 & 5.4 & N \nl 
4136 & Sc(r) & H$^1$ & 445 & 832 & 	2.4 & 4.7 & S,N \nl 
4144 & Scd & H$^1$ & 265 & 704 & 	6.5 & 7.1 & I \nl 
4178 (IC~3042) & SBc(s) & H$^1$ & 355 & 512 & 	13.8 & 6.0 & I \nl 
4183 & Scd & H$^1$ & 931 & 576 & 	4.1 & 7.4 & I \nl 
4190 (VV~104) & Sm & & 231 & 704 & 	6.5 & 2.0 & I,DN \nl 
4192 (M~98) & Sb & T2$^1$ & -140 & 640 & 	12.4 & 11.6 & S \nl 
4278 & E1 & L1.9$^1$ & 630 & 704 & 	12.2 & 4.8 & N \nl 
4293 & Sa & L2$^1$ & 933 & 768 & 	11.6 & 6.6 & N,I \nl 
4294 & SBc(s) & & 352 & 704 & 		22.8 & 3.8 & S \nl
4299 & Sd(s) & & 232 & 640 & 		23.5 & 2.0 & I,DN \nl 
4373 & E(4,2) & & 3444 & 896* & 	8.0 & 4.0 & N \nl 
4389 & SBc(s)pec & & 717 & 896 & 	20.3 & 3.1 & N,I \nl 
4395 & Sd & S1.8$^2$ & 317 & 320 & 	5.2 & 15.6 & N \nl 
4417 & S0(7) & & 843 & 448 & 		$<$ 0.1 & 4.0 & -- \nl 
4449 & Sm & H$^1$ & 207 & 1024 & 	107.3 & 7.3 & I,DN \nl 
4490 (VV~30) & Scd~pec & H$^1$ & 570 & 768 & 	43.1 & 7.4 & N,I,DN \nl 
4559 & Sc(s) & H$^1$ & 810 & 576 & 	28.0 & 12.6 & I \nl 
4571 (IC~3588) & Sc(s) & & 343 & 704 & 	0.3 & 4.2 & I \nl 
4605 & Sc(s) & H$^1$ & 141 & 960 & 	17.6 & 6.8 & I,DN \nl 
4701 & Sbc(s) & & 724 & 1024 & 		14.4 & 3.3 & S \nl 
4786 & E3 & & 4647 & 704* & 		10.6 & 1.9 & N \nl
4826 (M~64) & Sab(s) & T2$^1$ & 413 & 320 & 	69.1 & 11.8 & S,DL \nl 
5055 (M~63) & Sbc(s) & T2$^1$ & 503 & 512 & 	45.3 & 14.9 & S,I,DL \nl 
5444 & E3 & & 3974 & 960* & 		4.9 & 2.8 & N \nl 
5474 (VV~344) & Scd(s)~pec & H$^1$ & 275 & 576 & 	5.6 & 5.7 & I,DN \nl 
5585 & Sd(s) & H$^1$ & 304 & 576 & 	3.2 & 6.8 & I,DN \nl 
5605 & Sc(rs) & & 3363 & 640* & 	6.4 & 1.9 & I \nl 
5641 & SBab & & 4467 & 576* & 		3.6 & 3.0 & N \nl 
5653 & Sc(s)~pec & & 3557 & 512* & 	50.4 & 2.0 & S \nl 
5908 & Sb & & 3312 & 384* & 		33.9 & 3.8 & S,DL \nl 
6207 & Sc(s) & H$^1$ & 846 & 832 & 	33.2 & 3.5 & I \nl 
IC~4710 & SBd(s) & & 700 & 448 & 	2.9 & 4.2 & N \nl 
6684 & SBa(s) & & 886 & 768 & 		14.1 & 4.7 & N \nl 
6699 & Sbc(s) & & 3509 & 704* & 	9.2 & 1.8 & N,S,I \nl 
6744 & Sbc(r) & & 833 & 832 & 		3.4 & 23.6 & N \nl 
6754 & Sbc(s) & & 3325 & 512* & 	15.4 & 2.2 & N,S,I \nl 
6808 & Sc(s) & & 3466 & 576* & 		33.0 & 1.8 & S,I \nl 
6822 (IC~4895) & Im & & -49 & 512 & 	14.9 & 18.3 & I \nl 
6876 & E3 & & 3971 & 384* & 		$<$ 0.1 & 3.3 & -- \nl 
6946 & Sc(s) & H$^1$ & 48 & 768 & 	109.5 & 13.6 & N,S,I \nl 
IC~5052 & Sd & & 307 & 832 & 		27.0 & 7.0 & I \nl 
7309 & Sc(rs) & & 3938 & 1024* & 	10.3 & 2.2 & N,S,I \nl 
\enddata
\tablecomments{Column~1: NGC No. and possible alternative names.
Columns~2 and 4: morphological type
and recession velocity, resp., as taken from the RSA 
(\cite{san87}). Column~3: type of nuclear activity, adopted
from \cite{ho97a} (1) or \cite{ver98} (2): 
H II nucleus (H), Seyfert nucleus (S), LINER (L or S3), and 
transition object (T). Numbers attached to the class letter designate the 
type; ``:'' and ``::'' denote uncertain and highly uncertain classifications.
Galaxies that appear in neither of these catalogs
have no entry. Column~5: integration time for the narrow band filter 
(F190N with $*$, F187N otherwise). Column~6: total \paa -flux in
the field of view. Column~7: ratio of optical diameter $D_{25}$ (taken
from the RC2) to the NIC3 field of view of 51\as .
Column~8: morphology of the excess emission in the 
\paa\ image, if present: nuclear (N), 
HII regions in a spiral arm structure (S), a ring (R), 
or isolated throughout the disk (I), diffuse nebulosities (DN), 
or dust lanes (DL).}
\end{deluxetable}


\newpage
\begin{thebibliography}{}
\bibitem[Athanassoula 1992]{ath92}
	Athanassoula, E. 1992, MNRAS 259, 345
\bibitem[B\"oker, Krabbe, \& Storey 1998]{boe98}
	B\"oker, Krabbe, A., \& Storey, J.W.V. 1998, \apjl, 498, L115
\bibitem[B\"oker, F\"orster-Schreiber, \& Genzel 1997]{boe97}
	B\"oker, T., F\"orster-Schreiber, N., \& Genzel, R. 1997, \aj, 114, 1883
\bibitem[Buta \& Combes 1996]{but96}
	Buta, R., \& Combes, F. 1996, Fund. of Cosmic Phys. 17, 95
\bibitem[Deharveng \ea\ 1994]{deh94}
	Deharveng, J. M., Sasseen, T. P., Buat, V., Bowyer, S., Lampton, M., \& Wu, X.
	1994, A\&A , 289, 715
\bibitem[de Vaucouleurs, de Vaucouleurs, \& Corwin 1976]{dev76}
	de Vaucouleurs, G., de Vaucouleurs, A., \& Corwin, H.G. 1976, 
	Second Reference Catalogue of Bright Galaxies (University of Texas
	Press, Austin)
\bibitem[Gallais \ea\ 1991]{gal91}
	Gallais, P., Rouan, D., Lacombe, F., Tiph\'ene, D., \& Vauglin, I.
	1991, A\&A , 243, 309
\bibitem[Goudfrooij 1998]{gou98}
	Goudfrooij, P. 1998, to appear in ``Star Formation in Early-type Galaxies'', 
	PASP Conf. Series
\bibitem[Handa \ea\ 1990]{han90}
	Handa, T., Nakai, N., Sofue, Y., \& Hayashi, M. 1990, PASJ, 42, 1
\bibitem[Heckman \& Leitherer 1997]{hec97}
	Heckman, T. M. \& Leitherer, C. 1997, \aj , 114, 69
\bibitem[Ho, Filippenko \& Sargent (1997a)]{ho97a}
	Ho, L. C., Filippenko, A. V., \& Sargent, W. L. W. 1997a, \apjs , 112, 315
\bibitem[Ho, Filippenko \& Sargent (1997b)]{ho97b}
	Ho, L. C., Filippenko, A. V., \& Sargent, W. L. W. 1997b, \apj , 487, 568
\bibitem[Ishizuki \ea\ 1990]{ish90}
	Ishizuki, S., Kawabe, R., Ishiguro, M., Okumura, S. K., Morita, K.-I.,
	Chikada, Y., \& Kasuga, T. 1990, Nature, 344, 224  
\bibitem[Kennicutt 1983]{ken83a}
	Kennicutt, R. C. 1983, \apj , 272, 54
\bibitem[Kennicutt \& Kent 1983]{ken83b}
	Kennicutt, R. C. \& Kent, S. M. 1983, \aj , 88, 1094
\bibitem[MacKenty \ea\ 1997]{mac97}
	MacKenty, J.W., \ea\ 1997, ``NICMOS Instrument Handbook'', Version~2.0
	(Baltimore: STScI)
\bibitem[Meurer \ea\ 1995]{meu95}
	Meurer, G. R., Heckman, T. M., Leitherer, C., Kinney, A., Robert, C., \&
	Garnett, D. R. 1995, \aj , 100, 2665
\bibitem[Peng \ea\ (1996)]{pen96}
	Peng, R., Zhou, S., Whiteoak, J. B., Lo, K. Y., 
	\& Sutton, E. C. 1996, \apj, 470, 821
\bibitem[Sandage \& Tammann 1987]{san87}
	Sandage, A. \& Tammann G. A. 1987, ``A revised Shapley-Ames Catalog
	of bright Galaxies'', 2nd ed.,
	Carnegie Institution of Washington Publication (Washington, D.C.)
\bibitem[Shlosman, Frank, \& Begelmann 1989]{shl89}
	Shlosman, I., Frank, J., \& Begelmann, M.C. 1989, \nat , 338, 45
\bibitem[Shlosman, Begelmann, \& Frank 1990]{shl90}
	Shlosman, I., Begelmann, M.C., \& Frank, J. 1990, \nat , 345, 679
\bibitem[van den Bosch \& Emsellem 1998]{vdb98}
	van den Bosch, F. C. \& Emsellem, E. 1998, \mnras, 298, 276
\bibitem[Voit 1997]{voi97}
	Voit, M. (Editor) 1997, ``HST Data Handbook, Vol.~I'', Version~3.0 (Baltimore: STScI)
\bibitem[Veron-Cetty \& Veron (1998)]{ver98}
	Veron-Cetty, M. P. \& Veron, P., 1998, ``Quasars and Active Galactic Nuclei''
	(8th Ed.), ESO Sci. Rep. 18, 1
\bibitem[Waller \ea\ 1997]{wal97}
	Waller, W. \ea\ 1997, \apj , 481, 169
\bibitem[Young \ea\ (1996)]{you96}
	Young, J. S., Allen, L., Kenney, J. D. P., Lesser, A., \& Rownd, B. 
	1996, \aj , 112, 1903
\end{thebibliography}
\end{document}